\documentclass[sort&compress,final]{aipproc}\layoutstyle{8x11single}
\usepackage{amsmath}\usepackage{bm}\usepackage{booktabs}
\begin{document}\title{QCD Sum Rules for Heavy-Meson Decay
Constants:\\Impact of Renormalization Scale and Scheme}
\author{Wolfgang Lucha}{address={Institute for High Energy
Physics, Austrian Academy of Sciences, Nikolsdorfergasse 18,
A-1050 Vienna, Austria}}\author{Dmitri
Melikhov}{address={Institute for High Energy Physics, Austrian
Academy of Sciences, Nikolsdorfergasse 18, A-1050 Vienna,
Austria},altaddress={D.~V.~Skobeltsyn Institute of Nuclear
Physics, M.~V.~Lomonosov Moscow State University, 119991, Moscow,
Russia}}\author{Silvano Simula}{address={INFN, Sezione di Roma
Tre, Via della Vasca Navale 84, I-00146, Roma, Italy}}

\begin{abstract}Within the realm of QCD sum rules, one of the most
important areas of application of this nonperturbative approach is
the prediction of the decay constants of heavy mesons. However, in
spite of the fact that, indisputably, the adopted techniques are,
of course, very similar, we encounter rather dissimilar
challenges, or obstacles, when extracting from two-point
correlators of appropriate heavy-light currents interpolating the
mesons, the characteristics of charmed mesons with different spin.
In view of this, it seems worthwhile to us to revisit this issue
for the case of charmed pseudoscalar mesons $D_{(s)}$ and vector
mesons~$D^*_{(s)}.$\end{abstract}

\maketitle

\section{Approaching Bound States of Strong Interactions by QCD Sum
Rules}Any description of physical systems bound by the strong
interactions that deserves to be attributed as \emph{reliable\/}
should be based on quantum chromodynamics (QCD, the quantum field
theory that governs the strong interactions) and should be of
\emph{non\/}-perturbative nature. One formalism that --- in
contrast to, for instance, lattice gauge theory --- offers the
prospect of providing analytical insights, namely, in form of
relations between features of hadrons and the parameters of QCD~is
realized by the technique of \emph{QCD sum rules\/} \cite{SVZ}.
Their formulation proceeds along a well-established~sequence of
steps:\begin{itemize}\item Define the correlation function of a
nonlocal product of operators (in particular, of appropriate quark
currents) that interpolate the hadron under study, i.e., have
nonvanishing matrix elements between vacuum and this hadron~state.
\item Evaluate this correlation function at the hadron level, by
inserting a complete set of states, and at the QCD~level, by
applying Wilson's operator product expansion (OPE) reshaping any
nonlocal product to a series of local operators, to obtain
perturbative contributions, represented by dispersion integrals of
spectral densities, and non-perturbative (NP) terms, labelled as
``power'' contributions, representing the ``vacuum condensates''
of the local~OPE operators.\item Get rid of subtraction terms left
behind by Cauchy's integral formula and suppress the effects of
hadron excitations and continuum, by performing a Borel
transformation from momentum to another variable, the Borel
parameter~$\tau.$\item Hide your ignorance about higher states by
postulating quark--hadron duality; thus assume that all
contributions~of hadronic excited and continuum states cancel
against those of perturbative QCD above effective thresholds
$s_{\rm eff}(\tau)$.\end{itemize}

\section{Decay Constants of Pseudoscalar and Vector Charmed
Mesons $\bm{D}^{\bm{(}\bm{*}\bm{)}}_{\bm{(}\bm{s}\bm{)}}$}Taking
advantage of the experimental knowledge \cite{PDG} of the masses
$M_{\rm P,V}$ of the mesons discussed, our goal is to perform
advanced \cite{LMSAUa,LMSAUb,LMSAUc,LMSAUd,LMSAUe,LMSETa,LMSETb,
LMSETc,LMSETd,LMSETe} extractions of the decay constants, $f_{\rm
P,V},$ of both pseudoscalar (P) \cite{LMSDa,LMSDb} and
vector~(V)~\cite{LMSD*}~charmed mesons (regarded as bound states
of a charmed quark $c$ of mass $m_c$ and, in the non-strange case,
of a light quark $q=d$ of mass $m_d$ or, in the strange case, of a
light quark $q=s$ of mass $m_s$) from the two-point correlation
functions~of~adequately chosen interpolating currents. As
indicated, by way of construction the QCD sum rules derived along
the lines sketched above are expressed, at QCD level, in terms of
spectral densities $\rho^{\rm(P,V)}(s,\mu)$ and non-perturbative
terms $\Pi_{\rm NP}^{\rm(P,V)}(\tau,\mu)$ at appropriate
renormalization scale $\mu.$ Terming the QCD side of such sum rule
as the \emph{dual\/} correlator, $\widetilde\Pi_{\rm
P,V}(\tau,s_{\rm eff}(\tau)),$ we refer to the characteristics
predicted by this sum rule for a ground-state meson as its
\emph{dual\/} mass and \emph{dual\/} decay~constant:

\newpage

\begin{itemize}\item For the charmed pseudoscalar mesons ${\rm
P}=D,D_s,$ we select as their interpolating operator the
pseudoscalar current $j_5(x)\equiv(m_c+m_q)\,\bar q(x)\,{\rm
i}\,\gamma_5\,c(x)$ to extract \cite{LMSDa,LMSDb} both $M_{\rm P}$
and decay constants $f_{\rm P},$ defined by $\langle0|j_5(0)|{\rm
P}\rangle=f_{\rm P}\,M_{\rm P}^2$:
\begin{align*}&f_{\rm P}^2\,M_{\rm P}^4\exp\!\left(-M_{\rm
P}^2\,\tau\right)=\int\limits_{(m_c+m_q)^2}^{s_{\rm
eff}(\tau)}{\rm d}s\exp(-s\,\tau)\,\rho^{\rm(P)}(s,\mu)+\Pi_{\rm
NP}^{\rm(P)}(\tau,\mu)\equiv\widetilde\Pi_{\rm P}(\tau,s_{\rm
eff}(\tau))\ ,\\&M_{\rm dual}^2(\tau)\equiv-\frac{{\rm d}}{{\rm
d}\tau}\log\widetilde\Pi_{\rm P}(\tau,s_{\rm eff}(\tau))\ ,\qquad
f_{\rm dual}^2(\tau)\equiv\frac{\exp\!\left(M_{\rm
P}^2\,\tau\right)}{M_{\rm P}^4}\,\widetilde\Pi_{\rm P}(\tau,s_{\rm
eff}(\tau))\ .\end{align*}\item For the charmed vector mesons
${\rm V}=D^*\!\!,D_s^*,$ we use as interpolating operator the
vector current $j_\mu(x)\equiv\bar q(x)\,\gamma_\mu\,c(x)$ to
obtain \cite{LMSD*} the masses $M_{\rm V}$ and decay constants
$f_{\rm V},$ defined by $\langle0|j_\mu(0)|{\rm
V}(p)\rangle=f_{\rm V}\,M_{\rm V}\,\varepsilon_\mu(p)$ from the
sum~rule\begin{align*}&f_{\rm V}^2\,M_{\rm V}^2\exp\!\left(-M_{\rm
V}^2\,\tau\right)=\int\limits_{(m_c+m_q)^2}^{s_{\rm
eff}(\tau)}{\rm d}s\exp(-s\,\tau)\,\rho^{\rm(V)}(s,\mu)+\Pi_{\rm
NP}^{\rm(V)}(\tau,\mu)\equiv\widetilde\Pi_{\rm V}(\tau,s_{\rm
eff}(\tau))\ ,\\&M_{\rm dual}^2(\tau)\equiv-\frac{{\rm d}}{{\rm
d}\tau}\log\widetilde\Pi_{\rm V}(\tau,s_{\rm eff}(\tau))\ ,\qquad
f_{\rm dual}^2(\tau)\equiv\frac{\exp\!\left(M_{\rm
V}^2\,\tau\right)}{M_{\rm V}^2}\,\widetilde\Pi_{\rm V}(\tau,s_{\rm
eff}(\tau))\ .\end{align*}\end{itemize}The spectral densities are
known to three-loop accuracy \cite{SDa,SDb}; the values of our OPE
parameters are listed in Table~\ref{Tab:NV}.

\vspace{1.428ex}\begin{table}[ht]\caption{Input values (in
$\overline{\rm MS}$ renormalization scheme) chosen for quark
masses, QCD coupling and the lowest-dimensional vacuum
condensates.}\label{Tab:NV}
\begin{tabular}{cc}\toprule OPE parameter&Numerical input
value\\\midrule
$\overline{m}_d(2\;\mbox{GeV})$&$(3.42\pm0.09)\;\mbox{MeV}$\\[.5ex]
$\overline{m}_s(2\;\mbox{GeV})$&$(93.8\pm2.4)\;\mbox{MeV}$\\[.5ex]
$\overline{m}_c(\overline{m}_c)$&$(1275\pm25)\;\mbox{MeV}$\\[.5ex]
$\alpha_{\rm s}(M_Z)$&$0.1184\pm0.0020$\\[1ex]$\langle\bar
qq\rangle(2\;\mbox{GeV})$&$-[(267\pm17)\;\mbox{MeV}]^3$\\[.5ex]
$\langle\bar ss\rangle(2\;\mbox{GeV})$&
$(0.8\pm0.3)\times\langle\bar qq\rangle(2\;\mbox{GeV})$\\[.5ex]
$\displaystyle\left\langle\frac{\alpha_{\rm
s}}{\pi}\,GG\right\rangle$& $(0.024\pm0.012)\;\mbox{GeV}^4$\\
\bottomrule\end{tabular}\end{table}

\section{Progressing Towards Improved Predictions of Hadron
Observables}For Borelized QCD sum rules, progress in the achieved
precision \cite{LMSAUa,LMSAUb,LMSAUc,LMSAUd,LMSAUe} may be
hampered by too conventional attitudes:\begin{enumerate}\item The
requirement of Borel stability is nothing but a reflection of
one's mere hope that the value of a hadronic~feature predicted by
a QCD sum rule at an extremum in the Borel parameter is a reliable
approximation to the actual~value, but may lead one astray, as
experience with the counterparts of such sum rules in quantum
mechanics shows \cite{LMSAUa,LMSAUb,LMSAUc,LMSAUd,LMSAUe}.\item
The probably very na\"ive but persistently defended belief that
the effective threshold does not know about the Borel parameter
\cite{LMSETa,LMSETb,LMSETc,LMSETd,LMSETe}, i.e., the assumption
that the effective threshold is constant, is just a result of not
knowing~better.\footnote{Apart from our enduring campaign
\cite{LMSETa,LMSETb,LMSETc,LMSETd,LMSETe} against such
oversimplifying point of view, a notable exception is an
investigation of the decay~constants of heavy--light mesons
reported in Ref.~\cite{Gelhausen}, which hiddenly makes use of an
implicit dependence of the continuum threshold on the Borel
parameter.}\end{enumerate}In view of this, we proposed to allow
for the easy-to-find Borel parameter dependence of the effective
threshold \cite{LMSETa,LMSETb,LMSETc,LMSETd,LMSETe}:
\begin{itemize}\item Determine the range of admissible Borel
parameters $\tau$ --- the ``working Borel window'' --- by the
requirement that, at the window's lower end, the contribution of
the ground state is sufficiently large and, at the window's upper
end, the contributions of the nonperturbative corrections are
still reasonably small. For our analysis,
this~yields~\cite{LMSDa,LMSDb,LMSD*}
$$0.1\;\mbox{GeV}^{-2}<\tau<0.5\;\mbox{GeV}^{-2}
\quad\mbox{for}\quad\mbox{$D,$ $D^*,$ $D_s^*$}\
,\qquad0.1\;\mbox{GeV}^{-2}<\tau<0.6\;\mbox{GeV}^{-2}
\quad\mbox{for}\quad\mbox{$D_s$}\ .$$\item To derive the
Borel-parameter dependence of the effective thresholds $s_{\rm
eff}(\tau),$ adopt the simple polynomial Ansatz\footnote{Note that
this Ansatz allows for or covers (for $n=0$) but also generalizes
the conventional prejudice that the effective threshold should be
constant.}$$s^{(n)}_{\rm eff}(\tau)=\sum_{j=0}^ns_j\,\tau^j\
,\qquad n=0,1,2,\dots\ ,$$and pin down its coefficients $s_j$ by
minimizing over a set of $N$ equidistant discrete points $\tau_i$
in the Borel window the squared difference of dual meson mass
squared $M^2_{\rm dual}(\tau_i)$ and experimentally measured meson
mass squared~$M_{\rm P,V}^2$
$$\chi^2\equiv\frac{1}{N}\sum_{i=1}^N\left[M^2_{\rm
dual}(\tau_i)-M_{\rm P,V}^2\right]^2\ ,\qquad N=1,2,3,\dots\
.$$\item Having played around with several toy sum rules in
quantum mechanics \cite{LMSAUa,LMSAUb,LMSAUc,LMSAUd,LMSAUe,LMSETa,
LMSETb,LMSETc,LMSETd,LMSETe}, feel entitled to interpret
the~spread of results for the polynomial degree $n=1,2,3$ as a
hint to the size of the \emph{intrinsic\/} error of a QCD sum-rule
finding.\end{itemize}

\section{Systematic Uncertainties from Renormalization Scheme and
Scale}

\subsection{Issue: optimization of the perturbative convergence of
OPE contributions to QCD sum rules}
\begin{figure}[hbt]\begin{tabular}{cc}
\includegraphics[width=.491\columnwidth]{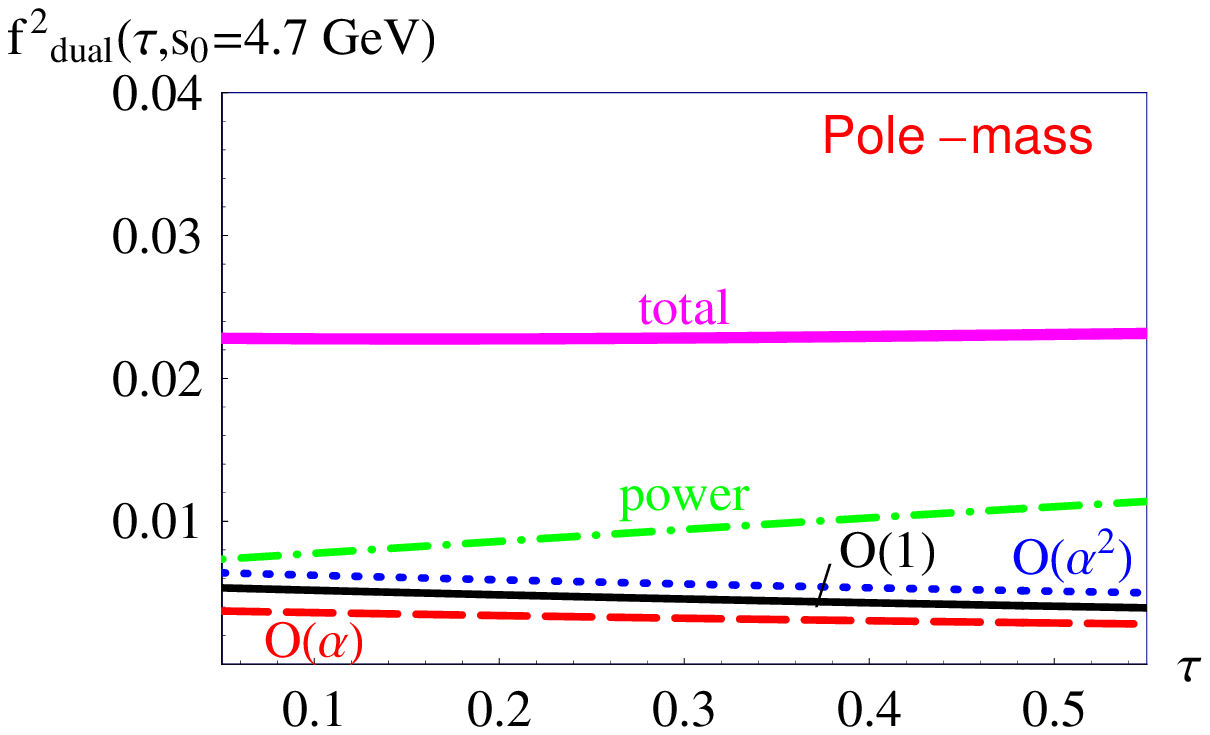}&
\includegraphics[width=.491\columnwidth]{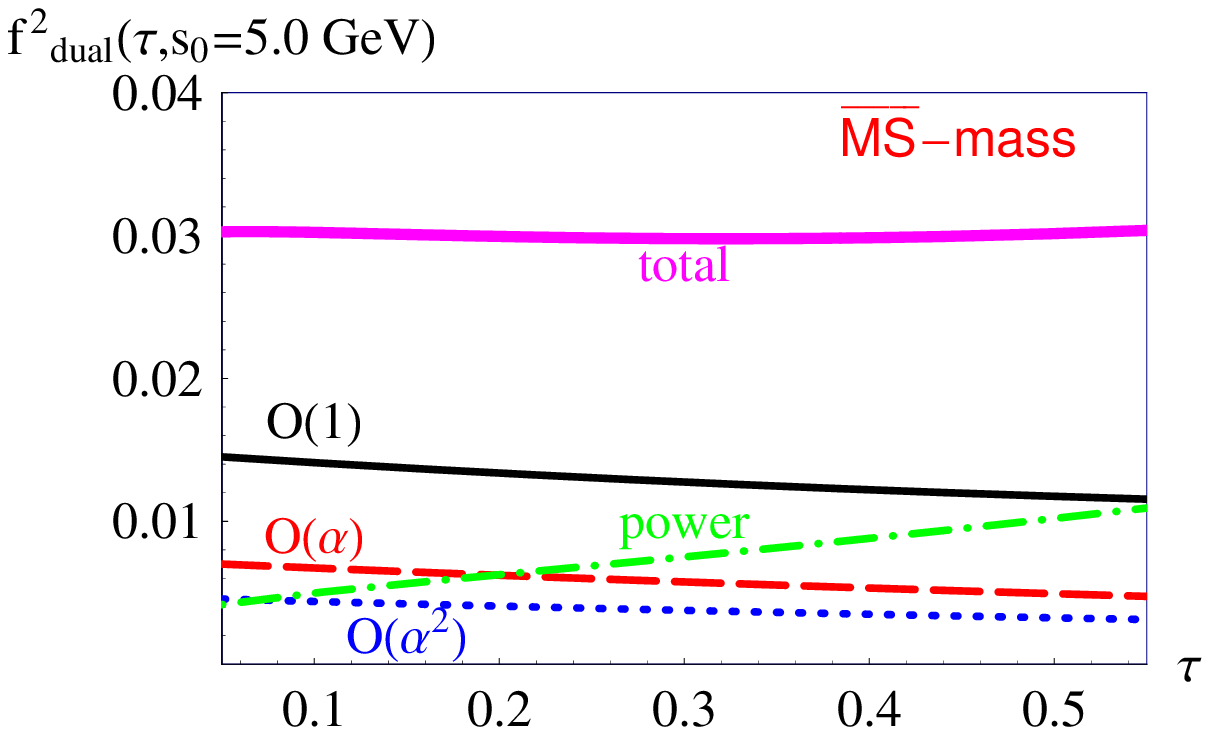}\\(a)&(b)
\label{Fig:Pscheme}\caption{Breakdown of the OPE contributions to
the dual decay constant $f_{D}(\tau)$ of the nonstrange charmed
pseudoscalar meson $D$ \cite{LMSDa,LMSDb}, extracted for a
\emph{fixed\/} threshold $s_0$ in either pole-mass renormalization
scheme (a) or \mbox{$\overline{\rm MS}$-mass}
renormalization~scheme~(b).}\end{tabular}\end{figure}

\begin{figure}[hbt]\begin{tabular}{cc}
\includegraphics[width=.491\columnwidth]{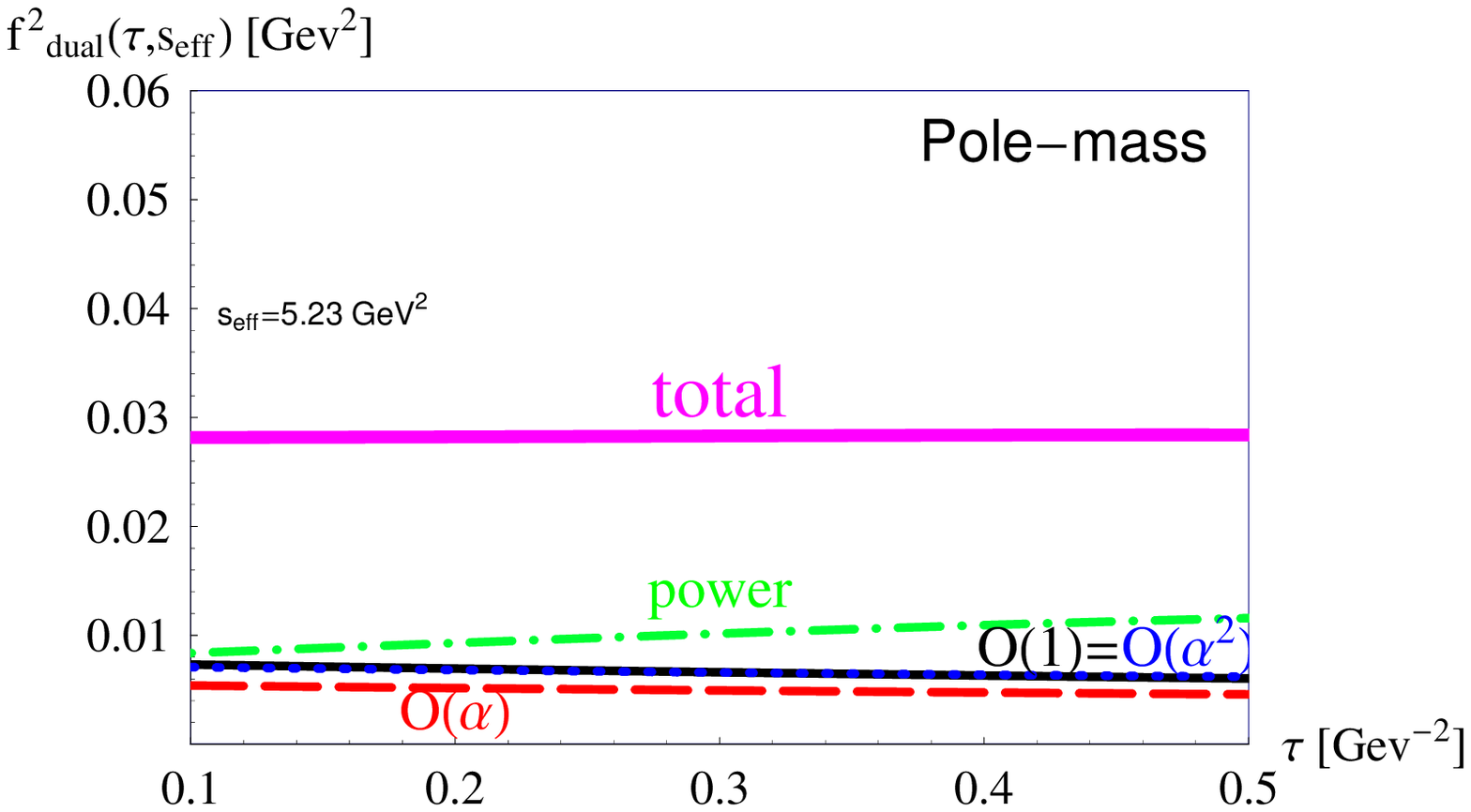}&
\includegraphics[width=.491\columnwidth]{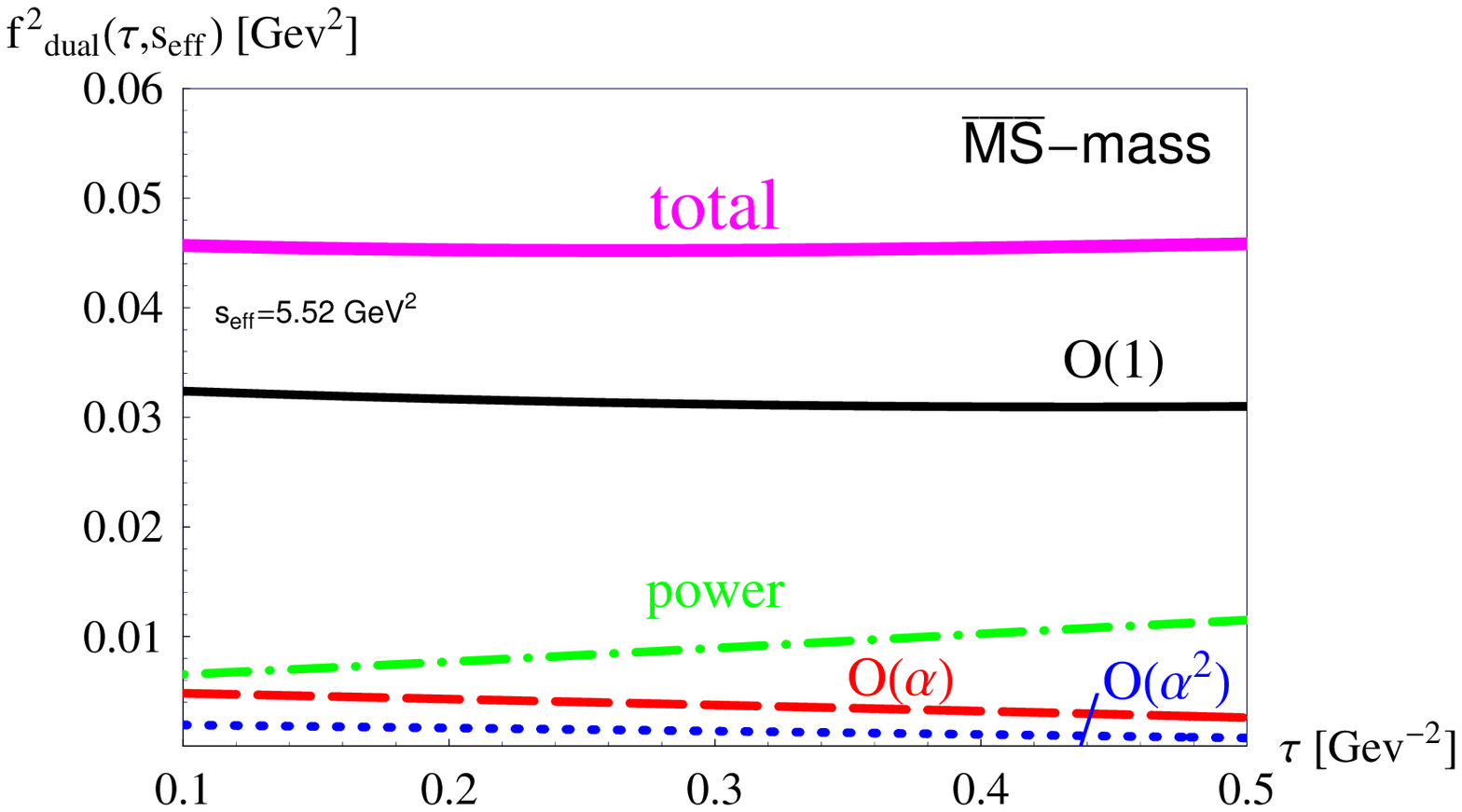}\\(a)&(b)
\label{Fig:Vscheme}\caption{Disentanglement of the OPE
contributions to the dual decay constant $f_{D^*}(\tau)$ of the
nonstrange charmed \emph{vector\/} meson $D^*$ \cite{LMSD*}
predicted, for a \emph{fixed\/} threshold $s_{\rm eff},$ by either
pole-mass renormalization scheme (a) or \mbox{$\overline{\rm
MS}$-mass} renormalization~scheme~(b).}\end{tabular}\end{figure}

Perturbation theory enables us
to derive the coefficient multiplying a given local operator in
some OPE in the form of a series in powers of the strong coupling,
$\alpha_{\rm s}(\mu).$ The one of the unit operator ends up in the
perturbative~spectral~density$$\rho(s,m_c,\mu)
=\rho_0(s,m_c)+\frac{\alpha_{\rm s}(\mu)}{\pi}\,\rho_1(s,m_c)
+\frac{\alpha_{\rm s}^2(\mu)}{\pi^2}\,\rho_2(s,m_c,\mu)+\cdots\
.$$ For the relative importance of the contributions both of
different order in $\alpha_{\rm s}$ and of power corrections to
predicted decay constants, the choice of the renormalization
scheme defining the $c$-quark mass, $m_c,$ makes a big difference:
although the central values are compatible within errors, the
comparisons shown, for the $D$ meson, in Fig.~\ref{Fig:Pscheme}
and, for the $D^*$ meson, in Fig.~\ref{Fig:Vscheme} assign a
greater credibility to results deriving from use of the
$\overline{\rm MS}$ running mass
$m_c=\overline{m}_c(\overline{m}_c)=(1275\pm25)\;\mbox{MeV}$ than
to those relying on the pole mass $m_c=\mathring{m}_c=1699\;{\rm
MeV},$ related to the former, via known expressions $r_1,r_2$
\cite{JL},~by $$\overline{m}_c(\mu)=\mathring{m}_c
\left(1+\frac{\alpha_{\rm s}(\mu)}{\pi}\,r_1+\frac{\alpha_{\rm
s}^2(\mu)}{\pi^2}\,r_2+\cdots\right).$$

\subsection{Issue: dependence of QCD sum-rule extractions of decay
constants on renormalization scale}
\begin{figure}[hbt]\begin{tabular}{cc}
\includegraphics[width=.491\columnwidth]{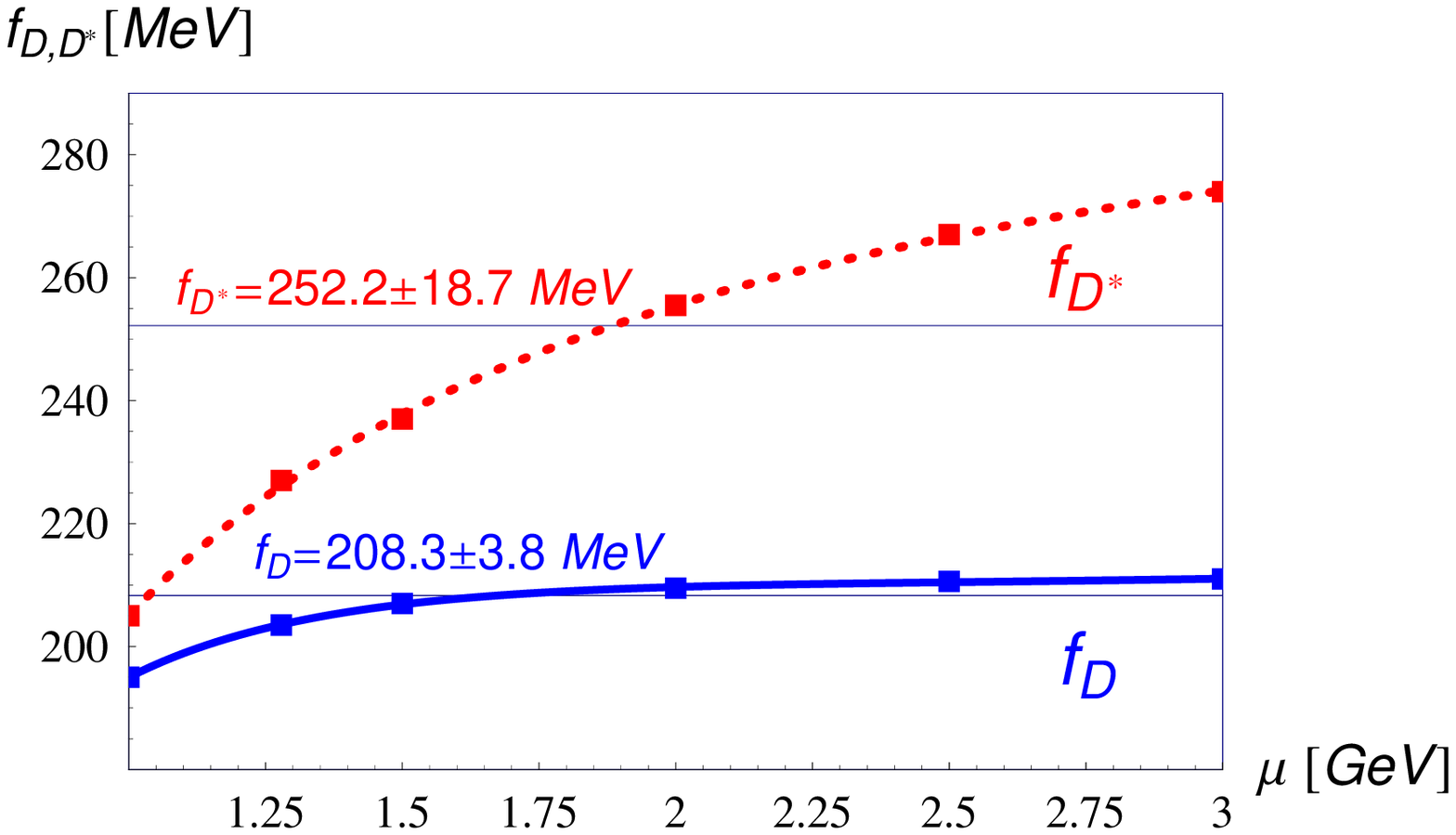}&
\includegraphics[width=.491\columnwidth]{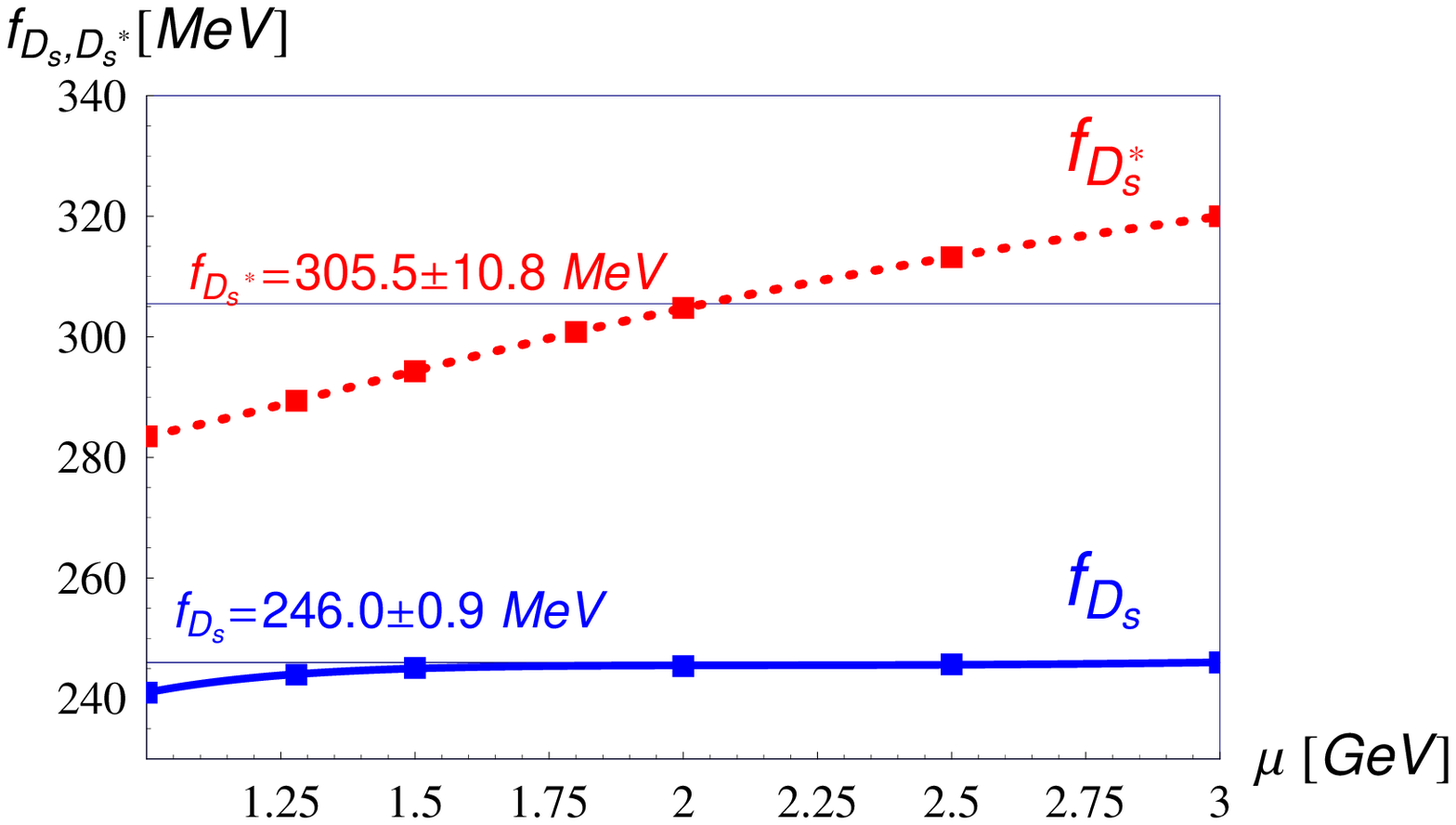}
\\(a)&(b)\caption{QCD sum-rule predictions for the unphysical
dependence of the dual decay constants $f_{D^{(*)}}$ of the
charmed nonstrange mesons $D,D^*$ (a) and of the dual decay
constants $f_{D_s^{(*)}}$ of the charmed strange mesons
$D_s,D_s^*$ (b) on the renormalization scale $\mu$~\cite{LMSD*}.}
\label{Fig:scale}\end{tabular}\end{figure}

Needless to recall,
\emph{physical observables\/} do not care about intermediate
technicalities such as renormalization scales: they do not depend
on any renormalization scales. Ideally, also theoretical
descriptions of such quantities should not do. Unfortunately,
within the formalism of QCD sum rules, for practical reasons
inevitable truncations, in their perturbative contributions, to
merely \emph{finite order\/} of the expansions in powers of the
coupling parameter and, in their nonperturbative power
corrections, to the relevant vacuum condensates of \emph{lowest
dimensions\/} induce an artificial unphysical dependence on
renormalization scales. In the case of decay constants, upon
introducing the average $\overline{\mu}$ of the renormalization
scale~$\mu$ by defining $f_{\rm dual}(\overline{\mu})=\langle
f_{\rm dual}(\mu)\rangle,$ such findings are approximately
reproduced by power series in the logarithm of
$\mu/\overline{\mu}$:\begin{equation}f_{D^{(*)}_{(s)}}(\mu)
=a\left(1+c_1\log\frac{\mu}{\overline{\mu}}
+c_2\log^2\frac{\mu}{\overline{\mu}}
+c_3\log^3\frac{\mu}{\overline{\mu}}\right).\label{Eq:RSD}
\end{equation}The numerical values of the scale average
$\overline{\mu}$ and of the parameters $a,c_1,c_2,c_3$ entering in
this expansion, emerging~from our QCD sum-rule extraction of the
decay constants of the charmed mesons $D,$ $D_s,$ $D^*,$ and
$D_s^*,$ are collected in Table~\ref{Tab:AS/SB}. The average scale
$\overline{\mu}$ is somewhat larger for vector mesons than for
pseudoscalar mesons. As evident from the numerical values of
primarily the coefficient $c_1,$ the sensitivity of the
charmed-meson decay constants to the renormalization scale $\mu,$
depicted in Fig.~\ref{Fig:scale}, is definitely more pronounced
for charmed vector mesons than for charmed pseudoscalar mesons.

\vspace{1.428ex}\begin{table}[ht]\caption{Outcomes \cite{LMSD*}
for average scale $\overline{\mu}$ and coefficients
$a,c_1,c_2,c_3$ in the parametrization (\ref{Eq:RSD}) of the
decay-constant renormalization scale behaviour.}\label{Tab:AS/SB}
\begin{tabular}{llllll}\toprule Charmed
meson&$\overline{\mu}\;\mbox{(GeV)}$&
\multicolumn{1}{c}{$a\;\mbox{(MeV)}$}&\multicolumn{1}{c}{$c_1$}&
\multicolumn{1}{c}{$c_2$}&\multicolumn{1}{c}{$c_3$}\\\midrule
$D$&1.62&$208.3$&$+0.06$&$-0.11$&$+0.08$\\
$D_s$&1.52&$246.0$&$+0.01$&$-0.03$&$+0.04$\\
$D^*$&1.84&$252.2$&$+0.233$&$-0.096$&$+0.17$\\
$D_s^*$&1.94&$305.5$&$+0.124$&$+0.014$&$-0.034$\\\bottomrule
\end{tabular}\end{table}

\section{Decay Constants of the Charmed Mesons: QCD Sum-Rule
Predictions}In this analysis, our goal was to take a fresh look,
from a common perspective, at our separate extractions
\cite{LMSDa,LMSDb,LMSD*} of the decay constants of both
pseudoscalar and vector charmed mesons. Qualitatively, we arrive
at the following~conclusions:\begin{itemize}\item When it comes to
the hierarchy of the perturbative and nonperturbative OPE
contributions, use of the $c$-quark mass defined by the
$\overline{\rm MS}$ renormalization scheme is the clear favourite
of both pseudoscalar and vector charmed~mesons.\item\emph{Central
values\/} of decay constants derived using the $\overline{\rm MS}$
$c$ mass are 30\% larger than those found from its pole~mass.\item
Unlike pseudoscalar mesons, vector mesons (viz., the
decay-constant errors) take notice of renormalization
scales.\end{itemize}

Quantitatively, our predictions for the decay constants of both
pseudoscalar \cite{LMSDa,LMSDb} and vector \cite{LMSD*} charmed
mesons (including the \emph{OPE-related\/} errors, caused by the
uncertainties of the parameter values entering as input to the
OPE,~and the \emph{systematic\/} errors, due to the inherently
limited accuracy of the QCD sum-rule approach) are summarized in
Table~\ref{Tab:DC}.

\vspace{1.428ex}\begin{table}[h]\caption{Reevaluation of the decay
constants of the charmed \emph{pseudoscalar\/} mesons $D_{(s)}$
\cite{LMSDa,LMSDb} and \emph{vector\/} mesons $D^*_{(s)}$
\cite{LMSD*} by our improved algorithm.}\label{Tab:DC}
\begin{tabular}{ll}\toprule Charmed meson&\multicolumn{1}{c}{Decay
constant $f_{D^{(*)}_{(s)}}\;\mbox{(MeV)}$}\\\midrule
$D$&$206.2\pm7.3_{\rm OPE}\pm5.1_{\rm syst}$\\[.5ex]
$D_s$&$245.3\pm15.7_{\rm OPE}\pm4.5_{\rm syst}$\\[.5ex]
$D^*$&$252.2\pm22.3_{\rm OPE}\pm4_{\rm syst}$\\[.5ex]
$D_s^*$&$305.5\pm26.8_{\rm OPE}\pm5_{\rm syst}$\\\bottomrule
\end{tabular}\end{table}

\bibliographystyle{aipproc}
\begin{thebibliography}{99}
\bibitem{SVZ}M.~A.~Shifman, A.~I.~Vainshtein, and V.~I.~Zakharov,
\emph{Nucl.~Phys.~B\/} \textbf{147} (1979) 385.
\bibitem{PDG}K.~A.~Olive et al.~(Particle Data Group),
\emph{Chin.~Phys.~C\/} \textbf{38} (2014) 090001.
\bibitem{LMSAUa}W.~Lucha, D.~Melikhov, and S.~Simula,
\emph{Phys.~Rev.~D\/} \textbf{76} (2007) 036002, arXiv:0705.0470
[hep-ph].
\bibitem{LMSAUb}W.~Lucha, D.~Melikhov, and S.~Simula,
\emph{Phys.~Lett.~B\/} \textbf{657} (2007) 148, arXiv:0709.1584
[hep-ph].
\bibitem{LMSAUc}W.~Lucha, D.~I.~Melikhov, and S.~Simula,
\emph{Phys.~Atom.~Nucl.\/}~\textbf{71} (2008) 1461.
\bibitem{LMSAUd}W.~Lucha, D.~Melikhov, and S.~Simula,
\emph{Phys.~Lett.~B\/} \textbf{671} (2009) 445, arXiv:0810.1920
[hep-ph].
\bibitem{LMSAUe}D.~Melikhov, \emph{Phys.~Lett.~B\/} \textbf{671}
(2009) 450, arXiv:0810.4497 [hep-ph].
\bibitem{LMSETa}W.~Lucha, D.~Melikhov, and S.~Simula,
\emph{Phys.~Rev.~D\/} \textbf{79} (2009) 096011, arXiv:0902.4202
[hep-ph].
\bibitem{LMSETb}W.~Lucha, D.~Melikhov, and S.~Simula,
\emph{J.~Phys.~G\/} \textbf{37} (2010) 035003, arXiv:0905.0963
[hep-ph].
\bibitem{LMSETc}W.~Lucha, D.~Melikhov, and S.~Simula,
\emph{Phys.~Lett.~B\/} \textbf{687} (2010) 48, arXiv: 0912.5017
[hep-ph].
\bibitem{LMSETd}W.~Lucha, D.~I.~Melikhov, and S.~Simula,
\emph{Phys.~Atom.~Nucl.\/}~\textbf{73} (2010) 1770,
arXiv:1003.1463 [hep-ph].
\bibitem{LMSETe} W.~Lucha, D.~Melikhov, H.~Sazdjian, and S.~Simula,
\emph{Phys.~Rev.~D\/} \textbf{80} (2009) 114028, arXiv:0910.3164
[hep-ph].
\bibitem{LMSDa}W.~Lucha, D.~Melikhov, and S.~Simula,
\emph{J.~Phys.~G\/} \textbf{38} (2011) 105002, arXiv:1008.2698
[hep-ph].
\bibitem{LMSDb}W.~Lucha, D.~Melikhov, and S.~Simula,
\emph{Phys.~Lett.~B\/} \textbf{701} (2011) 82, arXiv:1101.5986
[hep-ph].
\bibitem{LMSD*}W.~Lucha, D.~Melikhov, and S.~Simula,
\emph{Phys.~Lett.~B\/} \textbf{735} (2014) 12, arXiv:1404.0293
[hep-ph].
\bibitem{SDa}K.~G.~Chetyrkin and M.~Steinhauser,
\emph{Phys.~Lett.~B\/} \textbf{502} (2001) 104,
arXiv:hep-ph/0012002.
\bibitem{SDb}K.~G.~Chetyrkin and M.~Steinhauser,
\emph{Eur.~Phys.~J.~C\/} \textbf{21} (2001) 319,
arXiv:hep-ph/0108017.
\bibitem{Gelhausen}P.~Gelhausen, A.~Khodjamirian, A.~A.~Pivovarov,
and D.~Rosenthal, \emph{Phys.~Rev.~D\/} \textbf{88} (2013) 014015;
\textbf{89} (2014) 099901(E). 
\bibitem{JL}M.~Jamin and B.~O.~Lange, \emph{Phys.~Rev.~D\/}
\textbf{65} (2002) 056005, arXiv:hep-ph/0108135.
\end{thebibliography}
\end{document}